\documentstyle[prl,aps,twocolumn]{revtex}

\newcommand{\beq}{\begin{equation}}
\newcommand{\eeq}{\end{equation}}
\newcommand{\beqa}{\begin{eqnarray}}
\newcommand{\eeqa}{\end{eqnarray}}
\newcommand{\beqar}{\begin{eqnarray*}}
\newcommand{\eeqar}{\end{eqnarray*}}

\begin{document}

\title{ \bf
\Large Quantum limitations on superluminal propagation}

\author{ Yakir Aharonov$^{(a)}$,
Benni Reznik$^{(b)}$ 
and Ady Stern$^{(c)}$\\
{\ } \\
(a) {\it  \small School of Physics and Astronomy, Tel Aviv
University, Tel Aviv 69978, Israel, and }\\
{\it \small Department of Physics,
University of South Carolina, Columbia, SC 29208.}\\
(b) {\it \small Theoretical Division, T-6, MS B288,
Los Alamos National Laboratory, Los Alamos, NM, 87545}\\
(c) {\it \small Department of Condensed Matter Physics, Weizmann Institute of Science,
Rehovot 76100, Israel}
}

\maketitle

\begin{abstract}

  { Unstable systems such as media with inverted atomic population have
    been shown to allow the propagation of analytic wavepackets 
    with group velocity faster
    than that of light, without violating causality. 
    We illuminate the important role played by unstable modes in this
    propagation, and show that the quantum fluctuations of these
    modes, and their unitary time evolution, 
 impose severe restrictions on the observation of 
superluminal phenomena. }

\end{abstract}

A famous consequence of Einstein's special theory of relativity is the
principle of causality:  signals cannot travel faster than
light.
Nevertheless, it is known for some time,  that under certain conditions the
group velocity of an electromagnetic wavepacket can be arbitrarily large
and its energy positive,
but yet no conflict arises with causality:
the signal velocity remains always smaller than the velocity
of light in vacuum.
At least two  classes of such models allowing  ``causal''  superluminal
behavior,  are known: the first is closely related
to tunneling phenomena, the second arises for unstable systems of atoms 
under an inverted population condition\cite{review}.

While in the first case  the ``superluminal''
transmission of waves or particles
through a barrier is exponentially suppressed,
the second case  shows much more dramatic superluminal-like
effects. Wavepackets travel with a superluminal 
group velocity for unlimited distance, 
with negligible attenuation and dispersion. The second 
phenomenon arises due to instabilities in the initial 
state of the radiating system, for example, 
a scalar field initially in a
``false vacuum'' state  \cite{AKS}.
Recently, an optical experiment studying superluminal 
group velocities  for  waves in a medium
with inverted population was suggested
\cite{chiao,CB,CKK}.
While the discussion was limited so far to the semi-classical limit,
it was further suggested \cite{CKK},
that with a full fledged quantum-mechanical
treatment one would possibly obtain stable  tachyonic-like
quasiparticle excitations
in the inverted medium.

In this Letter we examine classical and quantum aspects of
superluminal group velocities in unstable systems. Classically, a peak
of an analytic wavepacket may travel from point $A$ to point $B$
faster than the speed of light, since its shape at $B$ can be fully
reconstructed from the part of the wave packet which is causally
connected to the point $B$.  We show that quantum mechanically,
analyticity of the wave packet is not the only condition needed for
such a reconstruction. Rather, the part of the wave packet which is
causally connected to the point $B$ must contain many photons or large
enough energy in the unstable modes.  This condition strongly
suppresses superluminal effects in the limit that the system contains
only a few photons.  The exponential suppression of superluminal
effects characterizes both the tunneling phenomena and the present
case of unstable systems.

To begin with we introduce a simple model which classically exhibits
tachyonlike motion.  Consider a real 
scalar field $\varphi(x,t)$ in one spatial dimension,  under the Hamiltonian
\beq
H = {1\over 2} \int dx\Bigl( \pi^2  +(\partial_x \varphi)^2
+\frac{2m^2}{\lambda}\cos{\sqrt{\lambda}\varphi}
\Bigr)
\label{hamiltonian}
\eeq Here, $\pi(x,t)=\dot\varphi(x,t)$ is the field conjugate to
$\varphi(x,t)$, $\lambda>0$ and the speed of light is put equal to
one. 
This Hamiltonian describes the
continuum limit of coupled pendula.  In the stable ground state
$\varphi(x)=\pi$.  We will be interested, however, in the dynamics of
$\varphi$ near the metastable state $\varphi(x)=0$.  For any finite time
interval of interest, we may take $\lambda$ sufficiently small such
that the potential term may be expanded,
$\cos{\sqrt{\lambda}\varphi}\approx 1-\frac{\lambda}{2} \varphi^2$.  From now
on we restrict ourselves to this expansion. Then, the equation of
motion for $\varphi$ becomes, \beq \Box \varphi - m^2 \varphi = 0
\label{eom}
\eeq
In term of the eigenmodes the solution is
\begin{eqnarray}
\varphi(x,t)= {1\over \sqrt{2\pi}}\int dk~ e^{ikx}\varphi_k(t) \\
\pi(x,t)= {1\over \sqrt{2\pi}}\int dk~  e^{ikx}\pi_k(t)
\end{eqnarray}
where reality conditions imply $\varphi_{-k}=\varphi_k^\dagger, ~~
\pi_{-k}=\pi_k^\dagger$. The time evolution of the modes is given by
\beqa
\varphi_k(t) &=& \varphi_{k0}\cos\omega_k t +
{\pi_{0k}\over \omega_k}\sin\omega_kt \nonumber \\
\pi_k(t) &=&\pi_{0k} \cos\omega_k t -
\omega_k  \varphi_{0k} \sin\omega_kt
\eeqa
where
\beq
\omega_k^2 = k^2 - m^2
\label{omega}
\eeq Notice that for $|k|>m$, $\omega_k$ is real, and $\varphi_k,\pi_k$
oscillate in time.  In the range $|k|<m$ $\omega_k$ is imaginary, and
$\varphi_k(t)$ and $\pi_k(t)$ are exponentially diverging.  The latter
modes are analogous to spontaneous emission in the optical model of an
inverted medium of two-level systems.  We will henceforth refer to the free
oscillatory modes and diverging modes as the ``normal'' and
``unstable'' modes, respectively. It is only near the point $\varphi\sim
\pi \sim 0$ that these modes exist as linearly independent solutions.
As $\varphi$ grows sufficiently, the instability is damped by the
non-linear $\lambda$-term in (\ref{hamiltonian}), but for the time of
interest to us, the latter can be neglected.  

We now turn to examine several features of the classical equation of
motion (\ref{eom}). First, we examine the propagation of a wavepacket
given by
\beq \varphi(x,t)=
\int_{{|k|>m}} {dk} \Bigl(
g_0(k)e^{ikx-i\omega_kt} + h.c.\Bigr) \eeq 
and $\pi(x,t)=\partial_t\varphi(x,t)$. We take
$g_0(k)$ to be non zero only in the range of {\it normal} modes
$m<|k|<k_{max}$, and centered
around $k=k_0$, with a width $\Delta k\ll k_0-m$. The spatial width of
this wavepacket is $1/\Delta k$, and we assume that at $t=0$ it is
spatially centered very far to the left of the origin, at $X_0\ll 0$.

This wavepacket propagates
with a group velocity, $v_g$, and a phase velocity, $v_p$, given by
\beq v_g = {1\over v_p} = {k_0 \over \sqrt{k_0^2-m^2}}
\label{vg}
\eeq
Since $v_g>1$, the motion of the center of the wave packet (group 
velocity) is
superluminal (tachyonlike), while the phase velocity
is always subluminal. As long as the dispersion is negligible, 
$\varphi(x,t)$ is 
(up to a time dependent phase factor) just $\varphi(x-v_gt, 0)$, i.e.,
the initial
wavepacket moving at velocity $v_g$. 

Second, we note that despite the superluminal group velocity, 
causality is maintained \cite{AKS,diener}. 
This is best seen through the 
Green's functions of Eq. (\ref{eom}).    
In terms of the homogeneous Green functions: $\varphi(x,t)=\int dx'
[ G(x-x',t)\varphi_0(x')+\tilde G(x-x',t) \pi_0(x')]$.
However, $\tilde G$ and $G$ vanish outside the light cone, i.e., for 
$x-x'>t$,  therefore 
{\it the value of $\varphi$ at point $x$ at time $t$ is fully 
determined by its value at points $x'$ that are causally connected to $x$.}

For further insight into this point, it is instructive to examine what
happens if the initial wavepacket is truncated at $x=0$, i.e.,
$\varphi_0(x)$ and $\pi_0(x)$ are replaced by $\varphi^T_0 =
{\tilde\theta}(x)\varphi_{0}(x)$ and  $\pi^T_0 =
{\tilde\theta}(x)\pi_{0}(x)$, respectively.  Here $\tilde\theta$ is a smoothed
step function. The length scale over which it is smoothed is assumed
smaller than all other length scales in the problem, and is kept
finite just to avoid an infinite $\partial_x\varphi$. Note that since
$\varphi_0(x)$ is centered at $X_0\ll 0$, the truncated wavepacket,
$\varphi^T_0(x)$ constitutes only the tail of the original one, and the
energy stored in the truncated wavepacket is just a small fraction of that
stored in the original one. The momentum representation of the
truncated wavepacket $\varphi_0^T(x)$ is $g_0^T(k)=\int
dk'\frac{1}{k-k'+i\eta} g_c(k')$, with $\eta$
an infinitesimal number.  The smoothing of the step function is
accounted for by introducing an upper cut--off to the $k$ integral.
While the original wavepacket was solely composed of normal modes, the
truncated one includes also unstable components.

The time evolution of the truncated wavepacket can be calculated using 
the exact expression for the Green function \cite{AKS,chiao,diener}.
There are three regimes: 
({\it a.}) For points $x<-t$, the causality of the Green function
dictates that the value of $\varphi_T(x,t)$ is not affected by the
existence of the truncated wavepacket at $x>0$. Thus, the field is
zero in this range.  ({\it b.}) For $-t<x<t$, however, the time
evolution of the truncated wavepacket is very different from that of
the original one: it is exponentially growing due to the contributions
of unstable components.  Since the amplitude of the normal modes
oscillates in time while the amplitude of the unstable modes grows
exponentially, for times $t\gg 1/m$ the wavepacket $\varphi^T(x,t)$ is
predominantly composed of unstable modes of very small wavevectors,
with an exponentially small contribution of normal modes.  ({\it c.})
Finally, for points $x>t$, the causality of the Green function
dictates that the value of $\varphi_T(x,t)$ does not depend on whether
or not it is truncated at the origin at $t=0$.  It is given, up to a
time dependent phase factor, by $\varphi(x-v_gt, 0)$. The propagation
of the superluminal peak is not affected by the truncation, and, for
long enough times, one finds at $X_0+v_gt$ a wavepacket of width
$1/\Delta k$, in which the field $\varphi$ oscillates with a
characteristic wavevector of $k_0$. Thus, at long times the field
$\varphi^T(x,t)$ shows an interesting behavior: it is predominantly
composed of very small wavevectors $|k|<m$ with just an exponentially
small contribution of wavevectors $|k|>m$. {\it However, in a region
  of spatial width $1/\Delta k$ around $X_0+v_gt$ it oscillates at the
  large wavevector $k_0>m$. The region $1/\Delta k$ can be made
  arbitrarily large, if $t$ is taken long enough.} These short-wavelength
oscillations of a long-wavelength wavepacket are very similar to the
superfast Fourier oscillations discussed in
\cite{aharonov,berry,reznik}.

The above discussion demonstrates that in order to maintain
consistency with causality, the mechanism which gives rise to
superluminal group velocities has to rely on the local information
stored in the tail.   As long as the true information that the
wave has been truncated has not arrived, the local
amplification extrapolates the peak as if this truncation does not
exist.  Classically, the possibility of extrapolating the full wave
from an infinitesimally small tail is, although surprising, possible.
As discussed in \cite{diener}, for an
analytic function, the wave can be reconstructed by means of a local
Taylor expansion, and the tachyonlike propagation can be viewed as 
an analytic continuation of the tail.

As we now show, the extrapolation of the full propagated wavepacket
from the truncated tail is made possible by the unstable modes, and
this is true independent of the details of the model. 
Since $g_0^T(k)$ is the Fourier transform
of a small tail, it is very small. The propagated wavepacket, after
truncation, is 
\beq 
\varphi^{T}(x ,t)= \int dk  \biggl(e^{ikx-i\omega_kt} 
g_0^T(k) + h.c.\biggr)
\eeq 
However while on the right hand side, the amplitudes
$g_0^T(k)$, originating from the truncated tail, are very small, 
on the left
side the magnitude of the tachyonic peak is not small, since the peak
is fully reconstructed from the tail. 
Since the amplitude of the stable modes, $\varphi_T(k),
$ for $ ~|k|>m$, is constant in time, this amplification can arise
only from the contribution of unstable modes growing exponentially with
time. It is then essential to have unstable modes for a full reconstruction 
of a superluminal peak from a truncated tail.

We now proceed to the quantum mechanical analog
of the classical superluminal system. Can a similar mechanism
of local amplification give rise to superluminal group
velocities? If so, it is conceivable
that by means of a process, analogous to the classical analytical
continuation,  this local information may be used to reconstruct
the wavepacket's peak.

The causality of the Hamiltonian (\ref{hamiltonian}) is manifested in
the quantum case by the statement that causally disconnected local
observables commute.  Again, this causality motivates us to study 
the quantum analog of the classical truncated wavepacket. In this
study, detailed below, we define an analog of a "false vacuum" state for
this model, on top of which we define a quantum state describing a
truncated wavepacket. Then, we show that if the truncated wavepacket is too
small (in a sense to be explained below), the initial quantum state
describing it is not orthogonal to the false vacuum state. In that case, the
vacuum quantum fluctuations in the fields $\varphi,\pi$ are, at $t=0$,
larger than the amplitude of the wavepacket, and one cannot
distinguish between the false vacuum state and the wavepacket state. As
we saw above, classically the small truncated
wavepacket becomes exponentially large in time, and one may naively
hope that quantum mechanically it eventually becomes larger than the
vacuum fluctuations. However, this cannot be the case, due to the
unitary quantum mechanical time evolution.  By the principle of
unitarity, the scalar product of two states, here the false vacuum and
truncated wavepacket states, is constant in time. Thus, if at $t=0$
the truncated wavepacket is not orthogonal to the false vacuum, it cannot be
distinguished from it at all later times. The exponential growth
of the tail is masked by an exponential growth in the quantum
fluctuations.

Turning to a more concrete discussion, we promote the functions
$\varphi(x,t)$, $\pi(x,t)$ to quantum operators.  From the standard
canonical commutation relations for the fields, $[\varphi(x,t),
\pi(x',t)] = i\delta(x-x')$, it follows as usual that for the normal
modes: $[\varphi_k, \pi_{k'}^\dagger]= [\varphi_k^\dagger,
\pi_{k'}]=i\delta(k-k')$. However, for the unstable modes the the
system does not admit the usual Fock space structure with an ordinary
particle interpretation.  For these modes ($|k|<m$), the Hamiltonian,
$\sum_k \pi_k^\dagger\pi_k -(m^2-k^2)\varphi_k^\dagger\varphi_k$, is
unbounded from bellow, and has a continuous and unbounded spectrum of
energies for each wave number $k$. Consequently, these modes cannot be
put in a ground state, and the fluctuations in the fields
$\varphi_k,\pi_k$ inevitably grow exponentially with time.

For the normal modes $|k|>m$ the vacuum state is obviously the ground
state of an harmonic oscillator of frequency $\sqrt{k^2-m^2}$. The
situation is less obvious for the unstable modes. On one hand, we
would like the quantum fluctuations in the fields $\varphi_k, \pi_k$ to
be as small as possible, but, on the other hand, stationary states of
the unstable modes necessarily have infinite fluctuations. Any state
of finite fluctuations is non--stationary, with the fluctuations
growing exponentially in time. Thus, we choose the initial state of
the unstable modes to be a non--stationary one, in which the
exponential growth of the field fluctuations is slowest. This state is
the ground state of an harmonic oscillator of frequency
$\sqrt{m^2-k^2}$. The vacuum state of the entire system is then a
direct product of the vacuum state for each mode $k$.

Next, we should find the quantum analog of the wave packet we analyzed
in the classical case.  Again, causality dictates that a
  measurement of the field $\varphi$ at a given point cannot be affected
  by its values in causally disconnected points. Thus, to analyze the
quantum analog of the classical super-luminal propagation, we need to
construct the quantum analog of the truncated wavepacket
$\varphi^T_0(x)$. We do so by shifting each mode $\varphi_k$ from its vacuum state,
in which $\langle\varphi_k\rangle =0$, to a coherent state, in which
$\langle\varphi_k\rangle =g^T_0(k)$:
\begin{equation}
|\Psi_0\rangle=\exp{\left[i\sum_k g^T_0(k){\hat\pi^\dagger}_k\right]}
\ |{\rm vac}\rangle
\label{qstate}
\end{equation}
where $|{\rm vac}\rangle$ is the  vacuum state.
Classically, as we saw above, the truncated wavepacket included all
the information needed to reconstruct the superluminal propagation of
$\varphi_0(x)$. {\it Quantum mechanically, however, this is true only if the
state (\ref{qstate}) is orthogonal to the vacuum state.} If that is
not the case, one cannot distinguish between the time evolution of the
vacuum state and that of (\ref{qstate}). 

The scalar product $\langle{\rm vac}|\Psi_0\rangle$ can be easily
calculated. It is, 
\begin{equation}
\langle{\rm vac}|\Psi_0\rangle=\exp{-\frac{1}{2\hbar}\int dk|\omega_k|
  (g_0^T(k))^2}
\label{scalar}
\end{equation}
It is useful to distinguish between the normal ($|k|>m$) and unstable
($|k|<m$) contributions to the $k$--integral in (\ref{scalar}). The
physical interpretation of the normal contributions is rather clear:
$g_0^T(k)$ is the amplitude of the oscillations of the mode $k$, and
$\omega(k)| (g_0^T(k))^2$ is the average number of photons in that
mode. The contribution of the unstable modes cannot be discussed in
terms of the photons. For these modes, the integrand
$\frac{1}{\hbar}|\omega(k)| (g_0^T(k))^2$ is just the ratio of the
kinetic energy stored in the $k$--mode to $\hbar\omega(k)$. {\it The quantum
  state containing the wavepacket, $|\Psi_0\rangle$, is then
  orthogonal to the vacuum state only if the truncated wavepacket
  contains many photons ($\gg 1$) of the normal modes, or large enough
  energy in the unstable modes.} This poses a new condition, which did
not exist in the classical theory, for the reconstruction of a
tachyonic peak from a truncated tail: the truncated tail cannot be too
small. If it is too small, the initial fluctuations in the field
$\varphi$, which grow exponentially with time, overcome the reconstructed
peak.

To exemplify the last point, let us evaluate the  
fluctuations of a local observable.
The field at a point has singular fluctuations, and one
needs to consider smeared operators like:
$
\bar \varphi_{(L,t)} \equiv \int f(x')\varphi(L-x',t) dx'.
$
where $f(x')$, is non-vanishing 
only in the spacetime volume $\Delta L$
around the point $L$, and is normalized  to $\int f dx=1$.  
Since 
$\langle \Psi_0 |\bar\varphi|
\Psi_0 \rangle$ is identical to the classically smeared field, the fluctuations are dominated 
by 
$
\langle \Psi_0 |\bar\varphi^2| \Psi_0 \rangle= \langle \bar\varphi^2
\rangle_N +  \langle \bar\varphi^2 \rangle_U
$
where the subscripts $N$ and $U$, stand for the contributions
of the normal and unstable modes. 
The contribution of the normal modes is always finite.
On the other hand, the unstable band yield an exponentially growing 
fluctuation:   
\beq
\langle \bar\varphi^2\rangle_U 
\approx
\hbar  I_0[2mT]  \stackrel{T \gg 1/m}{\rightarrow}
 \hbar {e^{2mT}\over \sqrt{4\pi mT}}
\label{uexp}
\eeq
where $I_0$ is zeroth order modified Bessel function.
For the observation of
superluminal propagation two conditions have to be satisfied: 
First $v_gT_{obs}\gg {1\over\Delta k}$ (where $T_{obs}$ is the time
at which the wavepacket is observed),
i.e., the point of observation
 should be  far outside the initial spread of the 
wavepacket. Second, we must distinguish between superluminal
propagation and propagation at the speed of light. This leads to:
$(v_g - 1) T_{obs} \gg {1\over \Delta k}$. These conditions require 
that $T_{min}m\gg 1$.
Eq. (\ref{uexp}) thus implies that 
for the signal amplitude to be larger than the amplitude of the
fluctuations at the observation time, the signal amplitude should ne
exponentially large. 

One may still question the generality of our result. For instance we
have assumed that initialy the unstable modes are in a direct product
state. Can a different initial state be made orthogonal to the vacuum
more easily?  We now argue that the obstacles pointed above in the way
to realizing quantum mechnical superluminal group velocities cannot be
circumvented.  Consider three initial classical tachyonlike
wavepackets, denoted by $R, L$ and $D$, with intial field and momenta
configurations $(\varphi_R, ~\pi_R)$, $(\varphi_L=\varphi_R,
~\pi_L=-\pi_R)$, and $(\varphi_D=-\varphi_R, \pi_D=\pi_R)$.  $R$ and
$L$ represent right and left moving wave packets with the same initial
$\varphi$, $D$ is right moving with negative $\varphi$.  Suppose that
initially the wavepackets are localized around $X_0\ll0$. Then after
some time, $R$ and $D$ form peaks (of opposite sign) at $X\gg0$ while
$L$ leaves at $X\gg 0$ only a small tail.  Again, we consider the
quantum states analogous to the classical truncated tails of the three
wavepackets, denoted by $R^T,L^T,D^T$, where the truncation is done at
$X=0$.  Classically, all three wavepackets can be reconstructed from
their tails.  {\it Quantum mechanically, however, it is not possible
  for all three quantum states to be mutually orthogonal}: $R^T$ and
$L^T$ are orthogonal if the uncertainty in the momentum in each state
is smaller than the difference between the momentum expectation value
of the two states: $\Delta\pi^T_{(L,R)}\ll
\langle\pi_R^T\rangle-\langle\pi_L^T\rangle$.  Similarly, $R$ and $D$
are orthogonal if $\Delta \varphi^T_{(R,D)} \ll
\langle\varphi_L^T\rangle-\langle\varphi_R^T\rangle$.  However, at the
tail, the right hand side of these two inequalities can be made
arbitrarily small, while the left hand side cannot be made arbitrarily
small, due to the uncertainty principle.  Thus, if the tails are small
enough, the classically
distinguishable wavepackets $R, L$ and $D$ cannot be mapped into three
mutually 
orthogonal quantum states.

In conclusion, we have shown that classical
superluminal -- like effects become incompatible with unitarity in the  
quantum mechanical limit, and are strongly suppressed.
Our argument strongly questions 
the possiblity that these systems may have 
tachyon-like quasiparticle excitations made of a small number of
photons \cite{CKK}.

{\bf Acknowledgment}
Y. A. acknowledges the support of  the Basic
Research Foundation,  grant 614/95,
administered by the Israel Academy of Sciences and Humanities.
A.S. acknowledges support of the US--Israel Bi-national Science Foundation
(95/250-1), the Minerva foundation and the V. Ehrlich career
development chair. We thank I. Ussishkin for constructive discussions.

\end{document}